\documentclass{llncs}
\usepackage[english]{babel}
\usepackage{stmaryrd}
\usepackage{graphics}
\usepackage{xspace}
\usepackage{dcolumn}
\usepackage{multicol}

\graphicspath{{./}{fig/}}

\usepackage[usenames]{color}

\newcommand{\dataloglb}{{Datalog$^{\mathrm{LB}}$}\xspace}

\newcolumntype{.}{D{.}{.}{-1}}

\usepackage{amsmath}
\usepackage{amssymb}
\usepackage{url}
\usepackage{fancyvrb}
\usepackage{multirow}
\usepackage{rotating}
\usepackage{arydshln}
\usepackage{graphicx}
\usepackage{subfig}

\title{Approximating Constraint Propagation in Datalog}

\author{
        Dario Campagna\inst{1}
\and
        Beata Sarna-Starosta\inst{2}
\and
        Tom Schrijvers\inst{3}
}

\institute{
Dept. of Mathematics and Computer Science, University of Perugia, Italy \\
\url{dario.campagna@dmi.unipg.it}
\and
LogicBlox Inc., Atlanta, Georgia, USA \\
\url{bss@logicblox.com}
\and
Dept. of Applied Mathematics and Computer Science, UGent, Belgium \\
\url{tom.schrijvers@ugent.be}}

\begin{document}

\maketitle

\begin{abstract}
We present a technique exploiting Datalog with aggregates to improve the 
performance of programs with arithmetic (in)equalities. Our approach employs 
a source-to-source program transformation which approximates the propagation 
technique from Constraint Programming.  The experimental evaluation of the 
approach shows good run time speed-ups on a range of non-recursive as well 
as recursive programs. Furthermore, our technique improves upon the previously 
reported in the literature constraint magic set transformation approach.
\end{abstract}

%==============================================================================%
\section{Introduction}\label{sec:intro}

{\em Datalog}~\cite{maier1988computing,abiteboul1995foundations} is a syntactic 
subset of Prolog introduced in the 1980s for database processing. By supporting
a limited, safe form of recursion, Datalog considerably extends the expressive 
power of traditional database query languages like SQL. At the same  
time, unlike Prolog, Datalog allows SQL's set-at-a-time evaluation. Also 
similarly to SQL, the programs in Datalog are guaranteed to terminate. Hence, 
extra-logical constructs such as Prolog's cut (`\texttt{!}') operator are not 
needed. 

After its original introduction as a smarter version of SQL, Datalog lost the
interest of researchers for a time, until recently re-gaining attention in
applications falling outside of the realm of traditional database reasoning,
which include: program
analysis~\cite{lam2005context,hajiyev2006codequest,hajiyev2006datalog},
networks~\cite{loo2006declarative,loo2005implementing}, security
protocols~\cite{li2003datalog}, knowledge representation~\cite{leone2006dlv},
robotics~\cite{ashley-rollman2007declarative} and
gaming~\cite{white2007scaling}. 
Our industrial partner, LogicBlox Inc.~\cite{logicblox}, uses a variant of
Datalog, called \dataloglb, as the basis for implementing decision automation 
and business planning systems.

Many of the above application domains rely on processing numerical data with
arithmetic operations, in Datalog available as built-in relations (predicates).
We focus in particular on built-in arithmetic (in)equality predicates ($>$, 
$<$, etc.) which we also refer to as \emph{(arithmetic) constraints}. The 
existing Datalog compilers do not exploit the constraining properties of the 
arithmetic predicates, but rather implement them as ordinary tests. As a 
result, evaluation of programs with arithmetic constraints follows the naive 
\textit{generate-and-test} approach, where ordinary predicates act as 
generators, and the entire search space that they produce is enumerated 
before the constraints can be applied to prune the candidate solutions. The 
research area of Constraint Programming (CP) offers approaches that prune the 
search space more eagerly, e.g., \emph{constrain-and-generate}, as well as the 
constraint implementation technique, called \textit{constraint propagators}, 
which allows to prune the domains of the variables involved in the constraints 
to narrow down the sets of candidate values even before the values are 
enumerated.

We adapt the CP constraint propagator technique to filter the individual 
generators in Datalog programs before they are used in the joins represented 
as arithmetic constraints. For this purpose we have developed an automatic 
program transformation framework in the \dataloglb system. Experimental 
evaluation shows that our technique enables good run-time improvements for 
a variety of test programs.

%==============================================================================%
\section{\dataloglb}
\label{sec:lb}

{\em LogicBlox} is a commercial Datalog-based platform for building 
enterprise-scale corporate planning and pricing applications. LogicBlox 
is currently used in several commercial decision automation applications, 
including retail supply-chain management~\cite{predictix} and software program 
analysis~\cite{bravenboer2009exception,bravenboer2009strictly,semmle}.
A typical LogicBlox application involves computational analyses that require 
aggregation across very large data sets combined with simulation and modeling 
techniques. The platform accommodates these features by means of its custom 
query language \dataloglb, a type-safe variant of Datalog, based on incremental 
evaluation, with trigger-like functionality and support for dynamic updates, 
declarative specification of functional dependencies, non-deterministic choice, 
stratified negation, meta-programming, and a wide range of extra-logical 
computations, including aggregation utilized by our optimization approach. 
In the following paragraphs we outline the main features of \dataloglb and 
the LogicBlox run-time system. A more exhaustive description of \dataloglb can
be found in~\cite{zook2009typed}. Readers familiar with Datalog may want to use 
this section as a reference when reading the remainder of the paper.

\paragraph{The \dataloglb Language. } 

\begin{figure}[t]
\begin{Verbatim}[frame=single,numbers=left]
digit(_) ->.
digit(d), val(d:v) -> uint[8](v), v<=9.

solution(i,a,m,s) -> digit(i), digit(a), digit(m), digit(s). 
solution(i,a,m,s) <-
   digit(i), val(i:vi),
   digit(a), val(a:va),
   digit(m), val(m:vm),
   digit(s), val(s:vs),
   vi != 0,
   vs != 0,
   vi != va, vi != vm, vi != vs,
   va != vm, va != vs,
   vm != vs,
   vi*(10*va+vm) = 100*vs+10*va+vm.
\end{Verbatim}
\caption{The \dataloglb encoding of the \texttt{I*AM=SAM} cryptarithmetic 
puzzle.}
\label{fig:iamsam}
\end{figure}

Figure~\ref{fig:iamsam} shows a \dataloglb encoding of the 
crypt\-a\-rith\-me\-tic puzzle \texttt{I*AM=SAM}, the goal of which is 
to find the assignment of digits to letters that satisfies the equation 
{\tt I*AM = SAM}. 

The basic programming construct in \dataloglb is the implication `{\tt <-}', 
denoting derivation rules of the form:
\begin{Verbatim}[xleftmargin=2cm]
Head <- Body.
\end{Verbatim}
where {\tt Head} and {\tt Body} are conjunctions of {\em atoms}. An atom 
can be either a predicate with variable or constant arguments, a comparison 
expression, an arithmetic expressions, or a negated atom. The above rule 
means that the atoms constituting {\tt Head} can be derived from the atoms 
constituting {\tt Body}. The example program in Figure~\ref{fig:iamsam} 
contains only one rule (lines 5-15), which derives the facts of the predicate 
{\tt solution} based on the facts of the predicates {\tt digit} and {\tt val}, 
and the constraints represented as comparisons and arithmetic expressions on 
their arguments.

\dataloglb extends Datalog with the notion of an {\em integrity constraint} 
of the form:
\begin{Verbatim}[xleftmargin=2cm]
Lhs -> Rhs.
\end{Verbatim}
Informally, the above constraint means that if {\tt Lhs} is true, then {\tt Rhs} 
must also be true, where {\tt Lhs} and {\tt Rhs} are conjunctions of atoms. The 
difference between a constraint and a rule is that a rule derives data for the 
atoms in its head, whereas a constraint checks that for the existing data 
matching its left-hand side, the right-hand side holds. The integrity 
constraints constitute the basis of \dataloglb's static type system, which 
guarantees at compile-time that certain kinds of constraints always hold for 
all possible instantiations of a given schema. Our approach uses integrity 
constraints to declare \emph{filter types} which allow to reduce the domains 
of predicates subjected to arithmetic constraints.

\dataloglb types are represented as unary predicates. Custom types 
may be defined using \emph{entity predicates}. For instance, in 
Figure~\ref{fig:iamsam}, the constraint in line 1 declares the entity 
predicate {\tt digit}. The constraint in line 4 is a \emph{type declaration} 
for the predicate {\tt solution}, which states that for every tuple 
{\tt solution(i,a,m,s)}, the arguments {\tt i}, {\tt a}, {\tt m}, and {\tt s} 
must be {\tt digit} entities. An entity predicate $P$ may be associated with a 
\emph{reference mode predicate}, which uniquely identifies each element in $P$ 
with a value of a primitive type, thus allowing to access the specific entity 
elements from user applications. For instance, line 2 of Figure~\ref{fig:iamsam} 
declares a reference mode predicate {\tt val}, which associates each entity 
element {\tt d} in {\tt digit} with {\tt v}, an 8-bit unsigned integer value 
no greater than 9, thus binding the {\tt digit} type to represent single-digit 
integers. The syntactic form {\tt val(d:v)} denotes the one-to-one functional 
relation between {\tt d} and {\tt v}, and is reserved for declaring reference 
mode predicates. The decision to express digits as entities is dictated by one 
of the mechanisms contributing to \dataloglb's termination guarantee, which 
restricts the use of primitive types as arguments to built-in predicates such 
as arithmetic operations.
%Type-based constraints can be checked at compile-time 
%by verifying that every rule that derives facts for {\tt solution} implies 
%the proper set membership for its arguments. 

The extra-logical operations supported by LogicBlox, including aggregation
computations, are represented by special-syntax rules of the form:
\begin{Verbatim}[xleftmargin=2cm,commandchars=\\\{\}]
result[x1,...,xn]=v <- \textit{Op}<<v=\textit{Method}>> \textit{Body}.
\end{Verbatim}
The head of the rule uses \dataloglb's shorthand notation for declaring 
functional dependencies: {\tt result[x1,...,xn]=v} declares the predicate 
{\tt result} to be a function from {\tt x1,...,xn} to {\tt v}. The notation 
also allows declaring singleton (constant) values: {\tt p[]=v} declares the 
predicate {\tt p} to be a singleton that contains only the value {\tt v}. The 
value can be retrieved through {\tt p[]}. The right-hand side of the above rule,
in addition to the conjunction of atoms in {\tt Body}, includes a directive 
which specifies the type of the operation to be performed (e.g., aggregation), 
and the particular method (e.g., finding the minimum value) to be used. For
instance, in Section~\ref{sec:nonrec} we show the following rule which finds
the lower bound for the {\tt val} predicate:
\begin{Verbatim}[xleftmargin=1cm]
lb_digit[]=n <- agg<<n=min(v)>> digit(d), val(d:v).
\end{Verbatim}
Above, {\tt agg} states that the rule computes an aggregation, and {\tt min}
names the specific operation to be applied to the values referenced by {\tt v}.
%, i.e., the integer values associated with the {\tt digit} type.

\paragraph{The LogicBlox Run-time System. }

\begin{figure}[t]
\centering
\scalebox{0.5}{\includegraphics{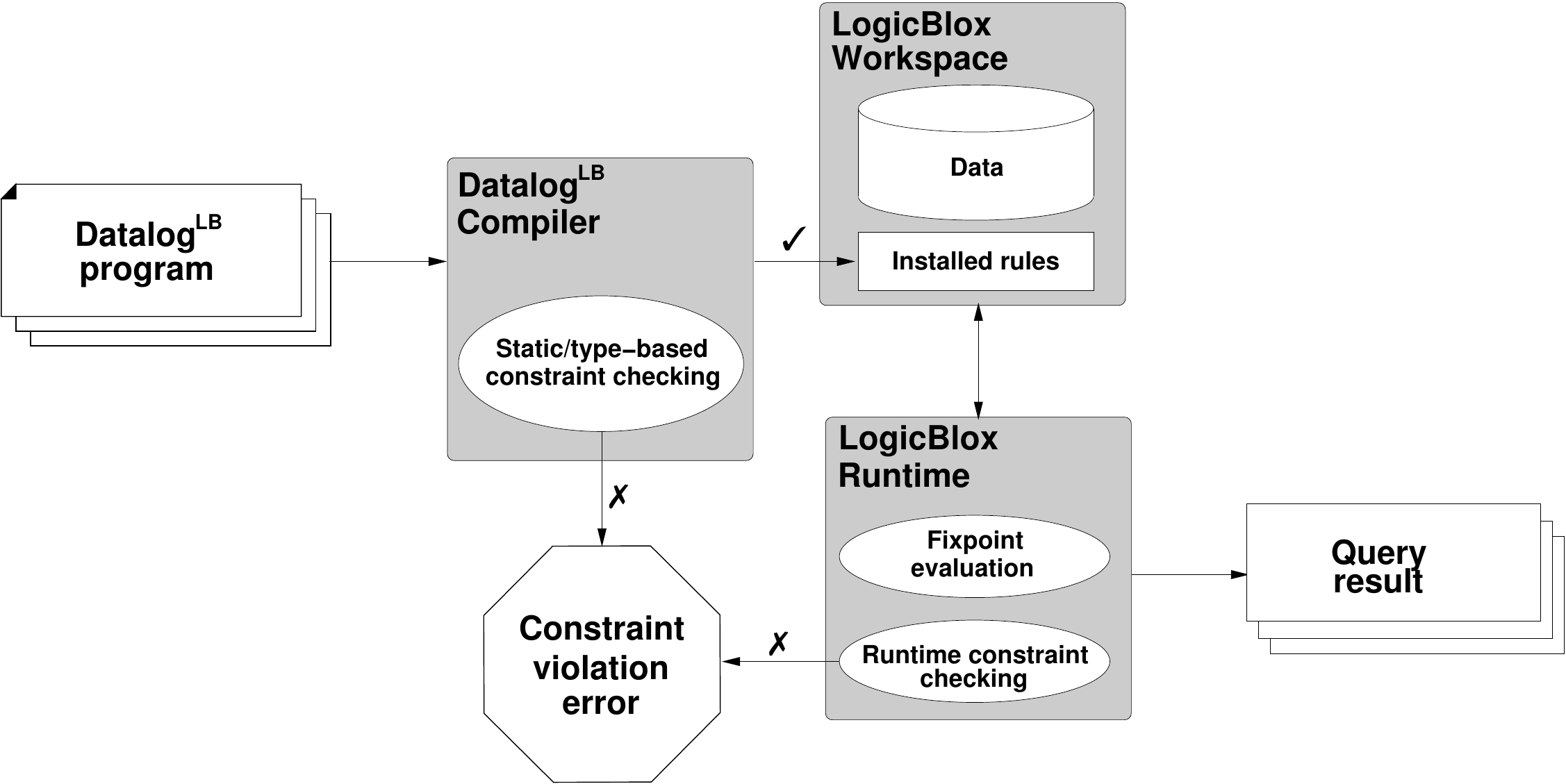}}
\caption{LogicBlox Architecture}
\label{fig:runtime}
\end{figure}

Figure~\ref{fig:runtime} illustrates the architecture of LogicBlox run time.
After compilation, discussed in more detail in Section~\ref{sec:impl}, the 
predicate definitions, rules, and constraints of the input \dataloglb program
are loaded into a designated data base instance called a \emph{workspace}. The 
workspace may be then accessed (queried and/or modified) from the LogicBlox 
API. Whenever a user application adds or removes facts from a predicate, the 
LogicBlox run-time engine incrementally re-evaluates the installed program 
rules until the workspace reaches a fixed point, i.e., no more facts can be 
derived by the rules. At the same time the engine checks the program's 
constraints, reporting any violations and restoring the workspace to a 
consistent state as needed.

%\begin{figure}[t]
%\centering
%\scalebox{0.5}{\includegraphics{fig/runtime_opt}}
%\caption{LogicBlox Architecture with bounds consistency optimization (in case we 
%want it somewhere, perhaps not right here)}
%\label{fig:runtime:opt}
%\end{figure}

%Figure~\ref{fig:runtime:opt} illustrates the architecture of LogicBlox run time
%with our optimization.

%==============================================================================%
\section{The Filter Predicates Transformation}\label{sec:transformation}

This section describes the details of our transformation, beginning with 
non-recursive programs, and then considering the impact of recursion.

%-------------------------------------------------------------------------------
\subsection{Non-Recursive Programs}
\label{sec:nonrec}
%
% Filter predicate form of the I AM SAM program.
%

Recall the \texttt{I*AM=SAM} program from the previous section. Our goal is 
to reduce the number of different candidate values that are used for producing 
answers. Thus, we exploit the equality constraint
\[ v_i *(10*v_a+v_m) = 100*v_s+10*v_a+v_m \] 
from the program rule to filter candidate values in the generator predicate 
\texttt{digit}.  Specifically, for each generator predicate atom appearing 
in the constraint, we consider the value generated by this atom in the context 
of the upper and lower bounds of the values produced by other generator atoms. 

For instance, for the generator atom \texttt{digit(i)}, the original constraint,
which is equivalent to the pair of inequalities:
\[\left\{\begin{array}{l}
v_i *(10*v_a+v_m) \leq 100*v_s+10*v_a+v_m\\
v_i *(10*v_a+v_m) \geq 100*v_s+10*v_a+v_m
  \end{array}\right.
\]
yields the pair of inequalities:
\[\left\{\begin{array}{l}
   v_i* (10*\,l_d\,+\,l_d\,) \leq 100*u_d+10*u_d+u_d \\
   v_i* (10*u_d+u_d) \geq 100*\,l_d\,+10*\,l_d\,+\,l_d 
  \end{array}\right.
\]
where $u_d$ and $l_d$ represent the upper and lower bound of the generator
predicate \texttt{digit}, respectively. We use these inequalities in the 
Datalog definition of the filter predicate for \texttt{digit(i)}:
\begin{Verbatim}[frame=single]
digit_filtered_i(i) <-
   digit(i),
   val(i:vi),
   lb_digit[]=t_1,
   ub_digit[]=t_2,
   vi* (10*t_1+t_1) <= 100*t_2+10*t_2+t_2,
   100*t_1+10*t_1+t_1 <= vi* (10*t_2+t_2).
\end{Verbatim}

\noindent
Similar filter predicates are generated for the remaining generator atoms.

The bounds for the generator predicates are computed in separate aggregate 
predicates, and reused in all filter predicates:

\begin{Verbatim}[frame=single]
lb_digit[]=n <- agg<<n=min(v)>> digit(d), val(d:v).
ub_digit[]=n <- agg<<n=max(v)>> digit(d), val(d:v).
\end{Verbatim}

In the last step of the transformation we replace the generator predicate atoms 
in the body of the \texttt{solution/4} rule by atoms representing corresponding 
filter predicates:

\begin{Verbatim}[frame=single]
solution(i,a,m,s) <-
   digit_filtered_i(i),
   digit_filtered_a(a),
   digit_filtered_m(m),
   digit_filtered_s(s),
   ... % rest of the original I AM SAM program
\end{Verbatim}

Our approach is inspired by the well-known \textit{bounds consistency} 
technique, in CP implemented by finite-domain constraint propagators. We 
simplify constraint propagation in two ways: (1) by computing filtered domains 
on the original domains rather than as a fixed point of the filtering process, 
and (2) by filtering only at the beginning of the evaluation rather than 
repeatedly after every enumeration step (in CP terminology known as 
\emph{labeling}).
%
%While we do miss out on potential speed-ups this way, the filter predicates
%approach remains a light-weight technique that is easily provided on top of 
%the existing Datalog implementations. As a consequence of this choice, (1) 
%we cannot encode unbounded fixpoint computations, and (2) computing and 
%storing many successively filtered tables for the same variable adds 
%considerable time and space overhead. Hence, our approach is a compromise 
%between anticipated speed-up and the overhead.
%
As a consequence of these simplifications, (1) we cannot encode unbounded 
fixpoint computations, and (2) computing and storing many successively filtered 
tables for the same variable adds considerable time and space overhead. 
Nevertheless, our approach yields a light-weight technique that is easily 
provided on top of the existing Datalog implementations, offering a satisfactory
compromise between the anticipated speed-up and the overhead.

%------------------------------------------------------------------------------
\subsection{Recursive Programs}

\begin{figure}[t]
\begin{Verbatim}[frame=single]
p(t,w) -> string(t), int[64](w).
s(t,w) -> string(t), int[64](w).

e(t,w) -> string(t), int[64](w).
e(t,w) <- p(t,w).
e(t,w) <- s(t,w),
          e(tp,wp),
          w - wp <= 100,
          w + wp >= 19500.
\end{Verbatim}
\caption{The \texttt{Engine} program.}\label{fig:rec}
\end{figure}

Recursion considerably complicates our transformation.  
%In this section we study the impact of the transformation on 
Consider the \texttt{Engine} program listed in Figure~\ref{fig:rec}. The program 
selects suitable engines for an engine yard. In the predicates \texttt{p(t,w)}, 
\texttt{s(t,w)} and \texttt{e(t,w)},  \texttt{t} corresponds to the engine 
type and \texttt{w} to the produced wattage. The predicate \texttt{p} 
represents the primary engines, and the predicate \texttt{s} represents 
the potentially spare engines. A suitable engine for the engine compound 
\texttt{e(t,w)} is either a primary engine, or a spare engine that can assist 
another engine in the compound. The difference in wattage between the assisting 
engine and the assistee should not exceed 100, and the total wattage of the 
compound should be no less than 19,500.

\begin{figure}[t]
\begin{Verbatim}[frame=single]
    e(t,w) <- p(t,w).                         s_filtered(t,w) <-
    e(t,w) <- s_filtered(t,w),                   s(t,w),
              e_filtered(tp,wp),                 w-ub_e[] <= 100,
              w-wp <= 100,                       19500 <= w+ub_e[].
              w+wp >= 19500.
                                              e_filtered(tp,wp) <-
                                                 e(tp,wp),
    lb_s[]=n <- agg<<n=min(v)>> s(_,v).          lb_s[]-wp <= 100,
    ub_s[]=n <- agg<<n=max(v)>> s(_,v).          19500 <= ub_s[]+wp.
    ub_e[]=n <- agg<<n=max(v)>> e(_,v). % ERROR 
\end{Verbatim}
\caption{Ill-formed \texttt{Engine} program after naive transformation.}
\label{fig:ill}
\end{figure}

The naive application of our technique yields the ill-formed program shown in
Figure~\ref{fig:ill}. The program involves recursion through aggregation: in 
order to compute the set of \texttt{e/2} we need to know the upper bound of 
\texttt{e/2}. Such recursion is not supported by \dataloglb{} (nor by any other 
LP system we are aware of).

Since it is not possible to effectively compute the \textit{exact} upper bound 
on \texttt{e/2}, we approximate it as the upper bound of the approximated upper 
bounds of the two rules defining \texttt{e/2}. For the first, non-recursive
rule, such an approximated (and exact) upper bound is \texttt{ub\_p[]}. A 
crudely approximated upper bound for the second, recursive rule, is 
\texttt{ub\_s[]}. Hence:
\begin{Verbatim}
      ub_e[]=n <- n=max(ub_p[],ub_s[]).
\end{Verbatim}
where 
\begin{Verbatim}
      ub_p[]=n <- agg<<n=max(v)>> p(_,v).
\end{Verbatim}

We may attempt to tighten the upper bound of the second rule, based on the 
observation that it is bounded from above by \texttt{ub\_e[]+100}:
\begin{Verbatim}
      ub_e[]=n <- n=max(ub_p[],min(ub_s[],ub_e[]+100)).
\end{Verbatim}
Alas, this step reintroduces recursion through aggregation. We eliminate 
it in the same way as before, by substituting the cruder approximation
derived earlier:
\begin{Verbatim}
   ub_e[]=n <- n=max(ub_p[],min(ub_s[],max(ub_p[],ub_s[])+100).
\end{Verbatim}
We further simplify the above expression by noticing that
\[ \forall x, y, c \in \mathbb{N}. \min(x,\max(y,x)+c) = x \]
This step brings us back to the first approximation, thus proving that 
the refinement attempt was unsuccessful. Nevertheless, as we show in 
Section~\ref{sec:eval}, our approximation is still quite effective at 
pruning the predicate domains and improving the performance of the programs.

\section{Implementation}\label{sec:impl}

%\paragraph{Implementation Language.} 
Most of the \dataloglb syntax is compatible with the term syntax of standard 
Prolog. The discrepancies in the particular notations, such as the functional 
dependency syntax, can be easily accommodated by simple processing steps. Hence, 
we chose Prolog (specifically, SWI-Prolog~\cite{swiprolog}) to implement the 
transformations of \dataloglb programs. Our analyzer consists of five Prolog 
modules, for the total of about 3,500 lines of Prolog code, including comments.

%-------------------------------------------------------------------------------
\subsection{LogicBlox/SWI-Prolog Interface} 

\begin{figure}[t]
\centering
\scalebox{0.5}{\includegraphics{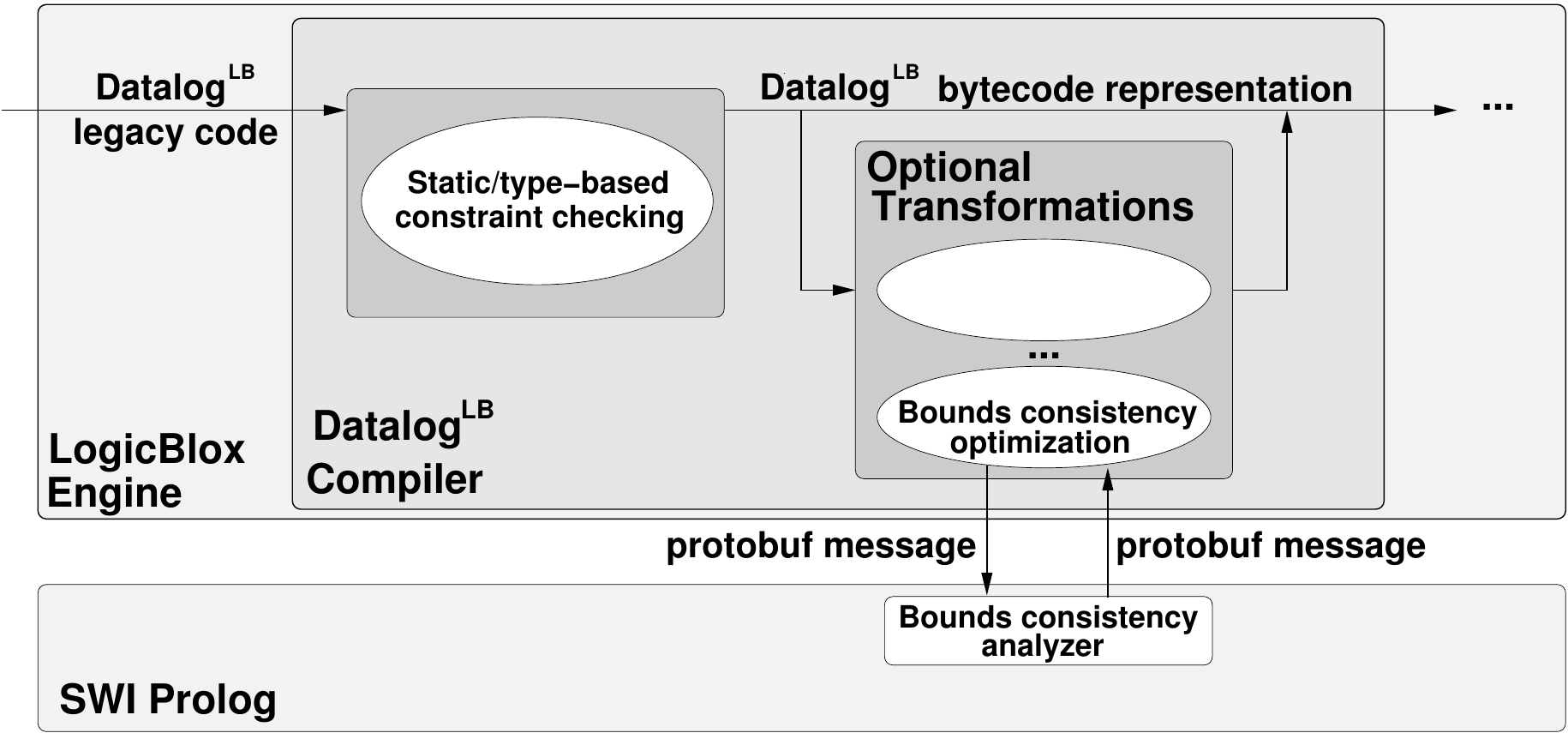}}
\caption{LogicBlox Compiler and its communication with SWI-Prolog}
\label{fig:compiler}
\end{figure}

Figure~\ref{fig:compiler} shows in more detail the LogicBlox compilation 
scheme and outlines the communication between the LogicBlox engine and the
SWI-Prolog analyzer. 

The LogicBlox compiler rewrites a source \dataloglb program into a core 
representation, which is then encoded using Google's protocol buffer (GPB) 
interface for further use by a number tools, including an interpreter.
GPB~\cite{protocolbuffers} is a platform-independent, extensible mechanism 
for serializing structured data. GPBs allow the programmers to determine how 
to structure their data by defining simple data structures (messages) in a 
dedicated specification language, and then to compile those data structures 
into the language and platform of their choice. As shown in 
Figure~\ref{fig:compiler}, the core program representation generated by 
the LogicBlox compiler is either passed directly to the subsequent phase of 
run-time processing, or subjected to one or more optional transformations 
aimed at optimizing the compiled code or collecting information to be used 
in further evaluation steps. This infrastructure makes the GPBs the medium 
of choice for interfacing LogicBlox with external analysis modules. In case 
of our application, the entire interface, including simplified message 
specification (discussed below), comprises five new modules, for the total 
of 1,800 lines of code.

The interface on the SWI-Prolog side is based on the system's native
GPB library~\cite{protobufs}, which we extended with support for necessary 
features of \dataloglb (e.g., recursively structured data), and optimized to 
linear run-time complexity. At this time our version of the library is 
available in a dedicated branch of the SWI-Prolog code repository.

The communication between LogicBlox and SWI-Prolog proceeds as follows. The 
output of the \dataloglb compiler is received by a new LogicBlox module which
extracts from it the information relevant to our analysis, encodes it as a 
collection of GPB messages, and opens a socket connection with SWI-Prolog. 
Once the connection is established, the messages are supplied to our analyzer.  
The analyzer decodes the messages into a program representation, applies the 
transformation, encodes the resulting program, and sends it back to the 
LogicBlox side, where another dedicated module retrieves the transformation 
results and updates the core representation of the program accordingly. 

We illustrate the use of GPBs on \dataloglb rule bodies. A rule body is 
a formula defined as an atom, a disjunction, a conjunction, or a negation. 
LogicBlox serializes and deserializes formulas with GBP messages of the 
following form: 
\begin{footnotesize}
\begin{Verbatim}[frame=single]
    message Formula {
        optional Atom atom = 1;
        optional Negation negation = 2;
        optional Conjunction conjunction = 3;
        optional Disjunction disjunction = 4;
    }
    message Conjunction { repeated Formula formula = 1; }
    ...
\end{Verbatim}
\end{footnotesize}
Note the mutually recursive nature of the \texttt{Formula} and
\texttt{Conjunction} definitions.

On the SWI-Prolog side, the messages are defined in \texttt{message/2} clauses:
\begin{footnotesize}
\begin{Verbatim}[frame=single]
protobufs:message(formula,Template) :-
    Template = [ optional(embedded(1,message(atom,_)),_)
               , optional(embedded(2,message(negation,_)),_)
               , optional(embedded(3,message(conjunction,_)),_)
               , optional(embedded(4,message(disjunction,_)),_)
               ].

protobufs:message(conjunction,Template) :-
   Template = [ repeated(1,embedded(_,message(formula,_))) ].
\end{Verbatim}
\end{footnotesize}
%The above uses two features we added to the SWI-Prolog GPB library:
%1) \texttt{message/2} for naming messages, essential for recursive messages, and 2)
The predicate \texttt{message/2}, which we added to the SWI-Prolog GPB library,
enables naming message templates. It is essential for recursive and 
repeated embedded messages. The \texttt{protobuf\_message/2} predicate 
serializes and deserializes messages to and from binary form, like the 
representation of the single-atom formula \texttt{digit(d)}.
\begin{footnotesize}
\begin{Verbatim}[frame=single]
?- protobuf_message(message(formula, 
        [ optional(embedded(1,
                     message(atom,
                         [ string(1,"digit"),
                           repeated(2,embedded([ ...
                                                 % Term for variable d
                                               ],term))
                         ])),present),
          optional(embedded(2,message(negation,_)),not_present),
          optional(embedded(3,message(conjunction,_)),not_present),
          optional(embedded(4,message(disjunction,_)),not_present)
        ]),Bytes).
\end{Verbatim}
\end{footnotesize}

%-------------------------------------------------------------------------------
\subsection{The Transformation} 

Given a representation of a \dataloglb program, our transformation processes in 
turn each of its rules. For every rule with one or more arithmetic constraints, 
it identifies the generator predicate atoms, exploits the constraints to produce
corresponding filter predicates, and replaces the generator atoms accordingly. 
It also extends the program with the definitions for the auxiliary predicates 
performing bounds computations.

Implementation of the code that generates the bounds-computing predicates 
turned out to be one of the more involved aspects of our project. 
The numerical data appearing in the arithmetic constraints pertinent to our
transformation is often represented as the values of the \dataloglb reference 
mode predicates where the keys are the entities produced by the predicates
serving as generators. To access these data, it is necessary to reconstruct
the chain of functional dependencies connecting each value with the appropriate
entity generator. For instance, to compute predicate bounds for the atom set:
\begin{Verbatim}[xleftmargin=.5cm]
	p(x), val_1(x:vx), q(y), val_2(y:vy), vx > vy
\end{Verbatim}
we need to reconstruct the chain connecting \verb|vx| with \verb|p| and 
\verb|vy| with \verb|q|. Additional complications arise when the reference mode 
predicates (and the corresponding generators) have non-unary keys, in which case 
the reconstructed dependencies are trees with the functional dependencies as 
nodes and the generators as leaves. 

%\paragraph{Keeping track of types. } 
\paragraph{}
As mentioned in Section~\ref{sec:lb}, \dataloglb's static type system relies
on the type information in the form of the integrity constraints. To ensure 
completeness of the type information in the transformed programs, we need to 
provide type declarations for the predicates generated by the analyzer (i.e., 
filter and aggregate predicates). It turns out that we can conveniently derive 
these directly from the original predicates, with no additional bookkeeping 
during the transformation.

%==============================================================================%
\section{Evaluation}
\label{sec:eval}
We now present the results of applying our transformation to a variety of 
programs. All experiments were performed using LogicBlox 3.7, on a machine 
with a 2.83 GHz Intel\textregistered\ Core\texttrademark\ 2 Quad CPU and 
4 GB of RAM, running Ubuntu 10.10 (Linux kernel 2.6.35-24-server). 
Additionally, we report the results obtained by the tabled top-down evaluation 
using XSB Prolog 3.3.1. For each experiment we show run times, in seconds, for 
the original programs, and the relative performance change after the 
transformation.

The changes required to accommodate \dataloglb programs in XSB are minimal
and mainly syntactic in nature: we omitted type declarations, replaced 
`\texttt{<-}' arrows with `\texttt{:-}', modified variable names to begin with 
capital letters, changed functional dependencies to ordinary arguments, and 
provided implementation for aggregates. The key feature to guarantee termination 
is tabling: all predicates are declared tabled.

% Changes made for XSB
%
% * dropping type declarations
% * replacing the arrows
% * dropping functional dependencies
% * aggregates -> reimplemented
% * tabling declaration

%-------------------------------------------------------------------------------
\subsection{Non-Recursive Programs}

%\subsubsection{Cryptarithmetic Puzzles} 
\paragraph{Cryptarithmetic Puzzles. } 

% \begin{table}[t]
% 	\centering
% 	\begin{tabular}{|r@{~=~}l|c|.|.|.|}
% 		\hline
% 		\multicolumn{2}{|c|}{\bf Puzzle}   & 
%                 \multicolumn{1}{|c|}{\bf Original} & 
%                 \multicolumn{1}{|c|}{\bf FP}       & 
%                 \multicolumn{1}{|c|}{\bf FP/Original \%} \\
% 		\hline
% 		\hline
% 		I * AM & SAM & 0.0088 & 0.0124 & 140.90 \\
% 		\hline
% 		BASE + BALL & GAMES & 0.8014 & 0.1124 & 14.02 \\
% 		\hline
% 		SEND + MORE & MONEY & 3.1033 & 0.3543 & 11.42 \\
% 		\hline
% 		BANJO + VIOLA & VIOLIN & 2.6724 & 0.3419 & 12.79 \\
% 		\hline
% 		SATURN + URANUS & PLANETS & 6.3935 & 0.9604 & 15.02 \\
% 		\hline
% 		SIX + SEVEN + SEVEN & TWENTY & 3.8952 & 1.0537 & 27.05 \\
% 		\hline
% 		DONALD + GERALD & ROBERT & 17.5421 & 18.4205 & 105.01 \\
% 		\hline
% 		BLACK + GREEN & ORANGE & 20.6298 & 2.4167 & 11.71 \\
% 		\hline
% 	\end{tabular}
% 	\caption{Benchmark results for cryptarithmetic puzzles.} % (LogicBlox Datalog version 3.7)
% 	\label{bench_cryptarithmetic}
% \end{table}

\begin{table}[t]
	\centering
%	\begin{tabular}{|r@{~=~}l||.@{\,s~}|.@{\,\%~}||.@{\,s~}|.@{\,\%~}|}
	\begin{tabular}{|r@{~=~}l||.@{\ sec.~}|.@{\,\%~}||.@{\ sec.~}|.@{\,\%~}|}
		\hline
		  			 \multicolumn{2}{|c||}{\multirow{2}{*}{\bf Puzzle}}   & 
                \multicolumn{2}{c||}{\bf \dataloglb} & 
                \multicolumn{2}{c|}{\bf XSB } \\
                %\cline{3-6}
                \multicolumn{2}{|c||}{} &
                \multicolumn{1}{c|}{\bf Original} &
                \multicolumn{1}{c||}{\bf FP} &
                \multicolumn{1}{c|}{\bf Original} &
                \multicolumn{1}{c|}{\bf FP} \\
		\hline
		\hline
		% I * AM & SAM & 0.0088 & 140.90 & 0.007 & 100.00 \\ %  0.0124 & 0.007 
		I * AM & SAM & 0.01 & 140.90 & 0.01 & 100.00 \\ %  0.0124 & 0.007 
		\hline
		% BASE + BALL & GAMES & 0.8014 & 14.02 & 6.46 & 13.00 \\ %  0.1124 & 0.84 XSB no reorder
		BASE + BALL & GAMES & 0.80 & 14.02 & 0.65 & 15.38 \\ %  0.1124 & 0.10
		\hline
		% SEND + MORE & MONEY & 3.1033 & 11.42 & 61.39 & 13.79 \\ %  0.3543 & 8.51 XSB no reorder
		SEND + MORE & MONEY & 3.10 & 11.42 & 2.60 & 11.92 \\ %  0.3543 & 0.31 
		\hline
		% BANJO + VIOLA & VIOLIN & 2.6724 & 12.79 & 61.53 & 13.88 \\ %  0.3419 & 8.54 XSB no reorder
		BANJO + VIOLA & VIOLIN & 2.67 & 12.79 & 2.73 & 12.09 \\ %  0.3419 & 0.33 
		\hline
		% SATURN + URANUS & PLANETS & 6.3935 & 15.02 & 630.13 & 14.53 \\ %  0.9604 & 91.58  XSB no reorder
		SATURN + URANUS & PLANETS & 6.39 & 15.02 & 7.70 & 12.60 \\ %  0.9604 & 0.97 
		\hline
		% SIX + SEVEN + SEVEN & TWENTY & 3.8952 & 27.05 & 645.98 & 24.09 \\ %  1.0537 & 155.65 XSB no reorder
		SIX + SEVEN + SEVEN & TWENTY & 3.90 & 27.05 & 8.75 & 25.26 \\ %  1.0537 & 2.21
		\hline
		% DONALD + GERALD & ROBERT & 17.5421 & 105.01 & TO & TO \\ %  18.4205 & XSB no reorder
		DONALD + GERALD & ROBERT & 17.54 & 105.01 & 17.20 & 107.62 \\ %  18.4205 & 18.51
		\hline
		% BLACK + GREEN & ORANGE & 20.6298 & 11.71 & TO & TO \\ %  2.4167 & XSB no reorder
		BLACK + GREEN & ORANGE & 20.63 & 11.71 & 19.99 & 52.53 \\ %  2.4167 & 10.5
		\hline
	\end{tabular}
	\caption{Benchmark results for cryptarithmetic puzzles.} % (LogicBlox Datalog version 3.7)
	\label{bench_cryptarithmetic}
\end{table}

Table~\ref{bench_cryptarithmetic} reports the evaluation run times for a set 
of cryptarithmetic puzzles building on the idea of the \texttt{I*AM=SAM} program 
from Section~\ref{sec:lb}. In almost all cases the transformation yields drastic
performance improvements (3$\times$ to 10$\times$). There are two exceptions. 
In case of the \texttt{I*AM=SAM} benchmark, the overhead of the auxiliary 
predicates introduced by the transformation dominates the extremely short run 
time of the original program. In case of the \texttt{DONALD+GERALD=ROBERT} 
puzzle, the transformed program prunes very few values from the initial domains, 
and consequently shows performance similar to that of the original program.

%We also consider an alternative backend: the tabled top-down evaluation
%mechanism of XSB Prolog 3.3.1. This backend exhibits a similar performance and
%benefits similarly from the transformation.
The XSB evaluation yields similar results both in terms of the original program
performance, and the benefits from the transformation.

% \begin{table}
% 	\centering
% 	\begin{tabular}{|r@{~=~}l|c|c|c|}
% 		\hline
% 		\multicolumn{2}{|c|}{\bf Puzzle} & {\bf Original} & {\bf FP} & {\bf FP/Original \%}\\
% 		\hline
% 		\hline
% 		I + AM & SAM & 0,0366 & 0,0455 & 124,32 \\
% 		\hline
% 		BASE + BALL & GAME & 12,2876 & 1,7407 & 14,17 \\
% 		\hline
% 		SEND + MORE & MONEY & 65,6277 & 7,1226 & 10,85 \\
% 		\hline
% 		BANJO + VIOLA & VIOLIN & 74,5806 & 7,1295 & 9,56 \\
% 		\hline
% 		SATURN + URANUS & PLANETS & 266,133 & 23,5132 & 8,83 \\
% 		\hline
% 		SIX + SEVEN + SEVEN & TWENTY & 102,971 & 21,4161 & 20,80 \\
% 		\hline
% 		DONALD + GERALD & ROBERT & 659,584 & 527,025 & 79,90 \\
% 		\hline
% 		BLACK + GREEN & ORANGE & 633,547 & 67,4304 & 10,64 \\
% 		\hline
% 	\end{tabular}
% 	\caption{Benchmark results for cryptarithmetic puzzles (LogicBlox Datalog version 3.6).}
% 	\label{bench_cryptarithmetic}
% \end{table}

%-------------------------------------------------------------------------------
%\subsubsection{The Production Problem}
\paragraph{The Production Problem. }
The \texttt{Production} program\footnote{We refer to
\url{http://users.ugent.be/~tschrijv/Datalog} for the source code.} models 
the mathematical programming problem of optimizing the profit from manufacturing
several types of products, subject to a set of constraints such as production 
costs and maximum number of items to be manufactured for each product type, or 
the availability of the factory line. From the technical point of view this 
program is interesting because it contains multi-key functional dependencies 
that drive the filter predicates. Another non-standard feature is the use of 
the aggregates for computing the optimized profit. 

Table~\ref{bench_production_multi_key} reports the results of evaluating the 
original and transformed program with four data sets differing in the range 
of the generator predicate indicating the number of tons of products being 
manufactured. Clearly, for LogicBlox evaluation, the transformation has no 
significant effect on the program for the small tons ranges, but enables a 
lot of pruning, and thus considerable performance improvement, when the tons 
ranges are large. On XSB the effects of the transformation are more uniform 
across the different data sets, with slightly better performance improvements 
for the larger tons ranges.

%
%\begin{table}[t]
%	\centering
%	\begin{tabular}{|c|.|.|.|}
%		\hline
%		\multicolumn{1}{|c|}{\bf Tons range} & 
%                \multicolumn{1}{|c|}{\bf Original} & 
%                \multicolumn{1}{|c|}{\bf FP} & 
%                \multicolumn{1}{|c|}{\bf FP/Original \%} \\
%		\hline
%         	\hline
%		[1,500] & 0.5989 & 0.6207 & 103.64 \\
%		\hline
%		[1,1000] & 2.8057 & 2.8269 & 100.75 \\
%		\hline
%		[1,2500] & 12.3609 & 5.0182 & 40.60 \\
%		\hline
%		[1,5000] & 13.7125 & 5.6597 & 41.27 \\
%		\hline
%	\end{tabular}
%	\caption{Benchmark results for production problem with multi-key functional dependency.} % (LogicBlox Datalog version 3.7).}
%	\label{bench_production_multi_key}
%\end{table}

\begin{table}[t]
	\centering
	\begin{tabular}{|c||.@{\ sec.~}|.@{\,\%~}||.@{\ sec.~}|.@{\,\%~}|}
		\hline
		\multicolumn{1}{|c||}{\multirow{2}{*}{\bf Tons range}} & 
                \multicolumn{2}{c||}{\bf \dataloglb} & 
                \multicolumn{2}{c|}{\bf XSB } \\
                % \cline{3-6}
                \multicolumn{1}{|c||}{} &
                \multicolumn{1}{c|}{\bf Original} &
                \multicolumn{1}{c||}{\bf FP} &
                \multicolumn{1}{c|}{\bf Original} &
                \multicolumn{1}{c|}{\bf FP} \\
		\hline
		\hline
		\hline
	%	[1,500] & 0.60 & 103.64 & 339.67\ {\rm sec.} & 99.86\ \% \\ % 0.6207 & 339.19
		[1,500] & 0.60 & 103.64 & 0.29 & 96.55 \\ % 0.6207 & 0.28
		\hline
		[1,1000] & 2.81 & 100.75 & 1.10 & 98.18 \\ % 2.8269 & 1.08
		\hline
		[1,2500] & 12.37 & 40.60 & 5.01 & 92.41 \\ % 5.0182 & 4.63
		\hline
		[1,5000] & 13.71 & 41.27 & 5.90 & 88.30 \\ % 5.6597 & 5.21
		\hline
        \end{tabular}
\caption{Benchmark results for the \texttt{Production} problem.}
%\caption{Benchmark results for \texttt{Production} problem with multi-key functional dependency.} % (LogicBlox Datalog version 3.7).}
\label{bench_production_multi_key}
\end{table}

%-------------------------------------------------------------------------------
\subsection{Recursive programs} 

%\subsubsection{The \texttt{engine} Program}
\paragraph{The Engine Program. }
To evaluate the effects of our transformation on the recursive \texttt{Engine} 
program from Figure~\ref{fig:rec}, we used four different data sets. Each data 
set defines the sets of couples produced by {\tt p/2} (denoted $P$ in the 
following), and {\tt s/2} (denoted $S$). Let 
$$\mathcal{T} = \{{\rm Steam\;engine},{\rm Internal\;combustion\;engine},{\rm
Gas\;Turbine}\}{\rm }$$ The four data sets define the sets $P$ and $S$ as
follows.
\begin{multicols}{2}
\begin{itemize}
	\item $Set_1$: $\left\{\begin{array}{rcl}
                          P & = & \mathcal{T} \times [1100,11500]\\
                          S & = & \mathcal{T} \times [1,10000]
                         \end{array}\right.$
	\item $Set_2$: $\left\{\begin{array}{rcl}
                          P & = & \mathcal{T} \times [500,5000]\}\\
                          S & = & \mathcal{T} \times [1,6000]\}
                         \end{array}\right.$
	\item $Set_3$: $\left\{\begin{array}{rcl}
                          P & = & \mathcal{T} \times  [500,16000]\}\\
                          S & = & \mathcal{T} \times  [1000,14000]\}
                         \end{array}\right.$
	\item $Set_4$: $\left\{\begin{array}{rcl}
                          P & = & \mathcal{T} \times  [10000,16000]\}\\
                          S & = & \mathcal{T} \times  [8,12000]\}
                         \end{array}\right.$
\end{itemize}
\end{multicols}
The results of the evaluation are shown in Table~\ref{bench_recursive}. There
is a visible correlation between the particular data set and the effects of the 
transformation. With little pruning possible, we see a modest speed-up or even 
a slow-down, whereas considerable pruning yields large performance improvements.
% \begin{table}
% 	\centering
% 	\begin{tabular}{|c|c|c|c|}
% 		\hline
% 		{\bf Data set} & {\bf Original} & {\bf FP} & {\bf FP/Original \%} \\
% 		\hline
% 		\hline
% 		$Set_1$ & 19,7701 & 3,7578 & 19,01 \\
% 		\hline
% 		$Set_2$ & 9,6216 & 0,4530 & 4,71 \\
% 		\hline
% 		$Set_3$ & 123,2 & 100,089 & 81,24 \\
% 		\hline
% 		$Set_4$ & 33,6853 & 31,8365 & 94,51 \\
% 		\hline
% 	\end{tabular}
% 	\caption{Benchmark results for recursive program in Figure~\ref{fig:rec} (LogicBlox Datalog version 3.6).}
% 	\label{bench_recursive}
% \end{table}

% \begin{table}[t]
% 	\centering
% 	\begin{tabular}{|c|.|.|.|}
% 		\hline
% 		{\bf Data set} & 
%                 \multicolumn{1}{|c|}{\bf Original} & 
%                 \multicolumn{1}{|c|}{\bf FP} & 
%                 \multicolumn{1}{|c|}{\bf FP/Original \%} \\
% 		\hline
% 		\hline
% 		$Set_1$ & 26.8660 & 5.7528 & 21.41 \\
% 		\hline
% 		$Set_2$ & 9.8253 & 0.4568 & 4.65 \\
% 		\hline
% 		$Set_3$ & 172.4680 & 145.1850 & 84.18 \\
% 		\hline
% 		$Set_4$ & 53.6065 & 56.1528 & 104.75 \\
% 		\hline
% 	\end{tabular}
% 	\caption{Benchmark results for recursive program in Figure~\ref{fig:rec}.} % (LogicBlox Datalog version 3.7).
%         \label{bench_recursive}
% \end{table}

\begin{table}[t]
	\centering
	\begin{tabular}{|c||.@{\ sec.~}|.@{\,\%~}||.@{\ sec.~}|.@{\,\%~}|}
		\hline
		  			 \multicolumn{1}{|c||}{\multirow{2}{*}{\bf Data set}}   & 
                \multicolumn{2}{c||}{\bf \dataloglb} & 
                \multicolumn{2}{c|}{\bf XSB } \\
                % \cline{3-6}
                \multicolumn{1}{|c||}{} &
                \multicolumn{1}{c|}{\bf Original} &
                \multicolumn{1}{c||}{\bf FP} &
                \multicolumn{1}{c|}{\bf Original} &
                \multicolumn{1}{c|}{\bf FP} \\
		\hline
		\hline
		$Set_1$ & 26.87 & 21.41 & 43.40 & 6.11 \\ % 5.7528 & 2.65  
		\hline
		$Set_2$ & 9.82 & 4.65 & 8.29 & 0.84 \\ % 0.4568 & 0,07
		\hline
		$Set_3$ & 172.47 & 84.18 & 119.82 & 64.91 \\ % 145.1850 & 77.78
		\hline
		$Set_4$ & 53.61 & 104.75 & 20.30 & 97.93 \\ % 56.1528 & 19.88
		\hline
	\end{tabular}
	\caption{Benchmark results for the \texttt{Engine} program.} % (LogicBlox Datalog version 3.7).
        \label{bench_recursive}
\end{table}

%-------------------------------------------------------------------------------
%\subsubsection{Multi-Legged Flights} 
\paragraph{Multi-Legged Flights Program. } 
The \texttt{Flights} program (Figure~\ref{fig:flights}) models multi-legged
flights and their travel distance.  More abstractly, it captures the transitive
closure of a directed weighted graph.
The \dataloglb encoding consists of the basic variant of the program, based on
that studied by Stuckey and Sudarshan~\cite{stuckey1994compiling}, together
with a sample query to compute all possible destinations no further than 10,000
miles from Sydney.

\begin{figure}[t]
\begin{Verbatim}[frame=single,numbers=left]
e(x,y,d) -> string(x), string(y), int[64](d).

f(x,y,d) -> string(x), string(y), int[64](d).
f(x,y,d) <- e(x,y,d), d >= 0.
f(x,y,d) <- e(x,z,d1), d1 >= 0, 
            f(z,y,d2), d2 >= 0, 
            d = d1 + d2, d <= 10000.

query(x,y,d) -> string(x), string(y), int[64](d).
query("Sydney",y,d) <- f("Sydney",y,d), d >= 0, d <= 10000.
\end{Verbatim}
\caption{The \dataloglb encoding of the \texttt{Flights} program.}
\label{fig:flights}
\end{figure}

Predicate {\tt e(x,y,d)} (line 1) denotes a flight leg, i.e., a direct
connection between cities {\tt x} and {\tt y} with the distance {\tt d}. 
The data of this predicate are given as facts.  The predicate {\tt f}
(lines 3-7) defines a multi-legged flight as the transitive closure of the
predicate {\tt e}. Since the second rule for {\tt f} contains recursion, to 
be expressible in \dataloglb, it needs to be bounded. Hence, we have added  
the constraint `{\tt d <= 10000}' (line 7), which is not present in the 
encoding of~\cite{stuckey1994compiling}. Lines 9-10 define the {\tt query} 
predicate.

It turns out that our transformation has no significant effect on the
performance  of the \texttt{Flights} program; it does not provide additional 
pruning. Fortunately, to our aid comes the \textit{constraint magic set 
transformation}~\cite{stuckey1994compiling}. Not only is the constraint magic 
set rewritten (CMR) variant of the program (Figure~\ref{fig:cmr}) faster than 
the original, but also it is amenable to our transformation. 

\begin{figure}[t]
\begin{Verbatim}[frame=single]
    answer_f(x,y,d) -> string(x), string(y), int[64](d).
    answer_f(x,y,d) <-
       x = "Sydney", f_a(x,y,d), d >= 0, d <= 10000.
    
    f_a(x,y,d) -> string(x), string(y), int[64](d).
    f_a(x,y,d) <-
       query_f_a(x,ld,ud), ld <= ud,
       e(x,y,d), d >= 0, d >= ld, d <= ud.
    f_a(x,y,d) <-
       query_f_a(x,ld,ud), ld <= ud,
       e(x,z,d1), d1 >= 0, 
       f_a(z,y,d2), d2 >= 0,
       d = d1 + d2, d >= ld, d <= ud.
    
    query_f_a(x,ld,ud) -> string(x), int[64](ld), int[64](ud).
    query_f_a("Sydney",0,10000).
    query_f_a(y,ld2,ud2) <-
       query_f_a(x,ld,ud), ld <= ud,
       e(x,y,d), d >= 0, 
       ud2 = ud - d, ld2 = max(ld-d,0).
    								 
    e(x,y,d) -> string(x), string(y), int[64](d).
\end{Verbatim}
\caption{Constraint magic rewritten variant of the \texttt{Flights} program.}\label{fig:cmr}
\end{figure}

%Now it does make sense to evaluate the program without (CMR) and with (CMR+FP)
%filter predicate transformation.  For that purpose we use a collection of 19
%different graphs, generated using three different methods.
Table~\ref{table:graphs} shows the results of evaluating the CMR variant of 
the \texttt{Flights} program without (CMR) and with (CMR+FP) filter predicate 
transformation for a collection of 19 different data graphs. The graphs, 
generated using three different methods, have the following structure:
\begin{itemize}	
\item there are six graphs with $n$ nodes, where each node has 
      [0,$\lfloor n/m\rfloor$] random bi-directional edges
      with random distance in [0,10000]. 
% An example of such a graph is depicted in Figure~\ref{sparse_graph}.

\item there are four graphs with $6\cdot n$ nodes consisting of 6 subgraphs with
      $n$ nodes. Within a subgraph, each node has $[0,m]$ bi-directional edges 
      with distance [0,7000]. Each subgraph is connected to $[0,o]$ other 
      subgraphs with distance [0,15000].
% Figure~\ref{subs_graph} consists of such a graph with $n=5$ and $m = o = 4$.

\item there are nine graphs with $n\cdot m$ nodes, consisting of $m$ complete 
      subgraphs of $n$ nodes that are not connected to one another.
% Figure~\ref{disc_graph} shows an example of such a disconnected graph with 
%$n = 8$ and $m = 2$.
\end{itemize}

%\begin{table}
%	\centering
%	\begin{tabular}{|*{3}{c|}}
%		\hline
%		{\bf Graph} & {\bf \# answers} & {\bf Percentage} \\
%		\hline
%		test\_sparse\_22-11 & 6 & 0,0051 \\		
%		test\_sparse\_33-11 & 26 & 0,0062 \\
%		test\_sparse\_36-12 & 187 & 0,0342 \\
%		test\_sparse\_40-20 & 950 & 1,2679 \\
%		test\_sparse\_50-15 & 1077 & 0,0864 \\
%		test\_sparse\_50-25 & 1667 & 1,6532 \\
%		\hline
%		test\_subs\_6-3-6 & 15 & 0,0083 \\
%		test\_subs\_5-2-3 & 264 & 0,1068 \\
%		test\_subs\_6-3-3 & 585 & 0,4194 \\
%		test\_subs\_8-4-6 & 1702 & 1,1383 \\
%		\hline
%		test\_disc\_9\_2 & 481 & 4,9876 \\
%		test\_disc\_9\_4 & 481 & 2,4938 \\
%		test\_disc\_9\_8 & 481 & 1,2469 \\
%		test\_disc\_8\_2 & 626 & 4,0862 \\
%		test\_disc\_8\_5 & 626 & 1,6345 \\
%		test\_disc\_8\_10 & 626 & 0,8172 \\
%		test\_disc\_10\_2 & 2638 & 6,3228 \\
%		test\_disc\_10\_6 & 2638 & 2,1076 \\
%		test\_disc\_10\_12 & 2638 & 1,0538 \\
%		\hline
%	\end{tabular}
%	\caption{Graphs used for flight program experiments.}
%	\label{graphs}
%\end{table}

% Figure~\ref{2D_answers} % and~\ref{3D_plot}, 
% how the runtime
% of the two programs changes with respect to the number of answers to the query
% considered. 
% Figure~\ref{3D_plot} shows how the run times vary with respect to
% both the number of answers and a percentage. The percentage has been computed
% as the number of answers to the query, divided by the number of tuples in the
% {\tt f}/3 relation of the original flight program. 
For the LogicBlox evaluation, Table~\ref{table:graphs} reports performance 
decrease for three transformed programs with corresponding original run times 
below 0.1s, and visible improvement for all other benchmarks. The speed-up 
varies roughly between 2$\times$ for the original programs with the shorter 
run times and 8$\times$ for those with longer run times. Interestingly, the
performance in XSB is very different. First, we observe that the run times for 
programs without the transformation are considerably shorter than in LogicBlox.
Furthermore, applying the transformation has no effect on the three programs 
with the shortest original run times, whereas it significantly slows down the
evaluation of all other programs. We attribute this negative effect to the 
ordering of constraints---imposed by our transformation when introducing
filter predicates---which forces overhead computations in the order-sensitive 
XSB.

\begin{table}[t]
	\centering
	\begin{tabular}{|@{\;Graph\;}l||.@{\ sec.~}|.@{\hspace{-.7cm}\%~}||.@{\ sec.~}|.@{\hspace{-.7cm}\%~}|}
		\hline
		\multicolumn{1}{|c||}{\multirow{2}{*}{\bf Graph}}   & 
                \multicolumn{2}{c||}{\bf \dataloglb} & 
                \multicolumn{2}{c|}{\bf XSB } \\
                % \cline{3-6}
                \multicolumn{1}{|c||}{} &
                \multicolumn{1}{c|}{\bf CMR} &
                \multicolumn{1}{c||}{\bf CMR+FP} &
                \multicolumn{1}{c|}{\bf CMR} &
                \multicolumn{1}{c|}{\bf CMR+FP} \\
		\hline
		\hline
		1 & 0.01 & 191.1 & 0.01 & 100.0 \\ % 0.001
		\hline
		2 & 0.03 & 162.0 & 0.01 & 100.0 \\ % 0.001
		\hline
		3 &  0.02 & 117.2 & 0.01 & 100.0 \\ % 0.001
		\hline
		\hline
		4 & 0.19 & 54.5 & 0.02 & 250.0 \\ % 0.05
		\hline
		5 & 4.47 & 21.2 & 0.51 & 468.6 \\ % 2.39
		\hline
		6 & 0.24 & 63.5 & 0.04 & 925.0 \\ % 0.37
		\hline
		7 & 0.76 & 41.7 & 0.12 & 2266.7 \\ % 2.72
		\hline
		8 & 2.91 & 22.1 & 0.31 & 442.8 \\ % 0.93
		\hline
		9 & 65.79 & 13.3 & 5.28 & 988.8 \\ % 52.21
		\hline
		10 & 5.76 & 42.0 & 1.26 & 504.8 \\ % 6.36
		\hline
		11 & 1.94 & 21.3 & 0.19 & 163.1 \\ % 0.31
		\hline
		12 & 2.40 & 38.0 & 0.39 & 2761.5 \\ % 10.77
		\hline
		13 & 2.83 & 22.1 & 0.29 & 320.7 \\ % 0.93
		\hline
		14 & 4.99 & 25.5 & 0.73 & 291.8 \\ % 2.13
		\hline
		15 & 66.93 & 13.0 & 5.14 & 1010.9 \\ % 51.96
		\hline
		16 & 1.92 & 22.9 & 0.17 & 170.6 \\ % 0.29
		\hline
		17 & 2.85 & 21.5 & 0.27 & 340.7 \\ % 0.92
		\hline
		18 & 1.92 & 21.4 & 0.16 & 181.2 \\ % 0.29
		\hline
		19 & 67.60 & 13.3 & 5.06 & 1030.0 \\ % 52.12
		\hline
	\end{tabular}
\caption{Benchmark results for the \texttt{Flights} program.}
\label{table:graphs}
\end{table}

%==============================================================================%
\section{Conclusion and Future Work} \label{sec:conclusion}

We presented a technique exploiting Datalog with aggregates to improve the 
performance of \dataloglb programs with arithmetic (in)equalities. Our approach 
employs a source-to-source program transformation that approximates the 
propagation technique from Constraint Programming. The experimental evaluation 
of the approach shows good run time speed-ups on a range of non-recursive as 
well as recursive programs. Furthermore, our technique improves upon the 
constraint magic set transformation approach proposed by Stuckey and Sudarshan.

In the future we plan to investigate ways to integrate finite domain solvers 
with the Datalog's semi-naive bottom-up evaluation mechanism to enable further 
benefits from constraint propagation. 
We would also like to compare our transformation-based approach to the tabled 
constraint programming approach proposed by Cui and Warren~\cite{bao}, applied 
to a finite domain constraint solver.

%==============================================================================%
\bibliographystyle{plain}
\bibliography{bib}

\begin{thebibliography}{10}

\bibitem{abiteboul1995foundations}
S.~Abiteboul, R.~Hull, and V.~Vianu.
\newblock {\em Foundations of Databases}.
\newblock Addison-Wesley, 1995.

\bibitem{ashley-rollman2007declarative}
M.~P. Ashley-Rollman, M.~De~Rosa, S.~S. Srinivasa, P.~Pillai, S.~C. Goldstein,
  and J.~D. Campbell.
\newblock {Declarative Programming for Modular Robots}.
\newblock In {\em Workshop on Self-Reconfigurable Robots/Systems and
  Applications at {IROS}}, 2007.

\bibitem{bravenboer2009exception}
M.~Bravenboer and Y.~Smaragdakis.
\newblock {Exception Analysis and Points-To Analysis: Better Together}.
\newblock In {\em ISSTA}, pages 1--12, 2009.

\bibitem{bravenboer2009strictly}
M.~Bravenboer and Y.~Smaragdakis.
\newblock {Strictly Declarative Specification of Sophisticated Points-To
  Analyses}.
\newblock In {\em OOPSLA}, pages 243--262, 2009.

\bibitem{bao}
B.~Cui and D.~S. Warren.
\newblock {A System for Tabled Constraint Logic Programming}.
\newblock In {\em CL}, pages 478--492, 2000.

\bibitem{protocolbuffers}
{Google's Protocol Buffers}.
\newblock \url{http://code.google.com/apis/protocolbuffers/}.

\bibitem{hajiyev2006datalog}
E.~Hajiyev, N.~Ongkingco, P.~Avgustinov, O.~de~Moor, D.~Sereni, J.~Tibble, and
  M.~Verbaere.
\newblock Datalog as a pointcut language in aspect-oriented programming.
\newblock In {\em {OOPSLA}}, 2006.

\bibitem{hajiyev2006codequest}
E.~Hajiyev, M.~Verbaere, and O.~de~Moor.
\newblock {CodeQuest: Scalable Source Code Queries with Datalog}.
\newblock In D.~Thomas, editor, {\em ECOOP}, 2006.

\bibitem{lam2005context}
M.~S. Lam, J.~Whaley, V.~B. Livshits, M.~C. Martin, D.~Avots, M.~Carbin, and
  C.~Unkel.
\newblock Context-sensitive program analysis as database queries.
\newblock In {\em PODS}, pages 1--12, 2005.

\bibitem{leone2006dlv}
N.~Leone, G.~Pfeifer, W.~Faber, T.~Eiter, G.~Gottlob, S.~Perri, and
  F.~Scarcello.
\newblock The {DLV} system for knowledge representation and reasoning.
\newblock {\em ACM Trans. Comput. Logic}, 7(3):499--562, 2006.

\bibitem{li2003datalog}
N.~Li and J.~C. Mitchell.
\newblock {Datalog with constraints: A foundation for trust management
  languages}.
\newblock In {\em PADL}, pages 58--73, 2003.

\bibitem{logicblox}
Logic{B}lox.
\newblock \url{http://logicblox.com/}.

\bibitem{loo2006declarative}
B.~Thau Loo, T.~Condie, M.~N. Garofalakis, D.~E. Gay, J.~M. Hellerstein,
  P.~Maniatis, R.~Ramakrishnan, T.~Roscoe, and I.~Stoica.
\newblock Declarative networking: language, execution and optimization.
\newblock In {\em Proceedings of the International Conference on Management of
  Data}, pages 97--108, 2006.

\bibitem{loo2005implementing}
B.~Thau Loo, T.~Condie, J.~M. Hellerstein, P.~Maniatis, T.~Roscoe, and
  I.~Stoica.
\newblock Implementing declarative overlays.
\newblock {\em SIGOPS Oper. Syst. Rev.}, 39(5), 2005.

\bibitem{maier1988computing}
D.~Maier and D.~S. Warren.
\newblock {\em Computing with Logic: Logic Programming with Prolog}.
\newblock Benjamin/Cummings, 1988.

\bibitem{predictix}
Predictix.
\newblock \url{http://www.predictix.com/}.

\bibitem{protobufs}
J.~Rosenwald.
\newblock {SWI-Prolog Google's Protocol Buffers library}.
\newblock \url{http://www.swi-prolog.org/pldoc/package/protobufs.html}.

\bibitem{semmle}
Semmle.
\newblock \url{http://semmle.com/}.

\bibitem{stuckey1994compiling}
P.~J. Stuckey and S.~Sudarshan.
\newblock Compiling query constraints (extended abstract).
\newblock In {\em PODS}, pages 56--67, 1994.

\bibitem{white2007scaling}
W.~White, A.~Demers, C.~Koch, J.~Gehrke, and R.~Rajagopalan.
\newblock Scaling games to epic proportions.
\newblock In {\em Proceedings of the ACM SIGMOD International Conference on
  Management of Data}, pages 31--42, 2007.

\bibitem{swiprolog}
J.~Wielemaker.
\newblock {SWI-Prolog 5.10 Reference Manual}.
\newblock \url{http://www.swi-prolog.org}, April 2010.

\bibitem{zook2009typed}
D.~Zook, E.~Pasalic, and B.~Sarna-Starosta.
\newblock Typed {D}atalog.
\newblock In {\em PADL}, pages 168--182, 2009.

\end{thebibliography}
%==============================================================================%
\appendix

% \begin{figure}
% 	\centering
% 	\subfloat[]{%[test\_sparse\_22-11.]{
% 		\label{sparse_graph}
% 		\includegraphics[scale=0.2]{test_sparse_22-11}
% 	}\\
% 	\subfloat[]{%[test\_subs\_5-2-3]{
% 		\label{subs_graph}
% 		\includegraphics[scale=0.2]{test_subs_5-2-3}
% 	}\\
% 	\subfloat[]{%[test\_disc\_8\_2]{
% 		\label{disc_graph}
% 		\includegraphics[scale=0.2]{test_disc_8_2}
% 	}
% 	\caption{Graphical representation of three test graphs (each undirected edge represents a couple of oppositely directed edges).}
% \end{figure}

% \begin{figure}
% 	\centering
% 	\includegraphics{2D_answers}
% 	\caption{Flight programs: run time with respect to number of answers.}
% 	\label{2D_answers}
% \end{figure}
% 
% \begin{figure}
% 	\centering
% 	\includegraphics{3D_plot}
% 	\caption{Flight programs: run time with respect to number of answers and percentage.}
% 	\label{3D_plot}
% \end{figure}
%==============================================================================%

\end{document}